\documentclass[twocolumn,tighten,times]{aastex631}
\usepackage{color}
\usepackage{multirow}
\usepackage{chngcntr}
\usepackage{mathtools}
\usepackage[utf8x]{inputenc}
\usepackage{perpage}
\MakePerPage{footnote}
\usepackage{lipsum}
\usepackage[graphicx]{realboxes}
\makeatletter

\newcommand{\Rmnum}[1]{\expandafter\@slowromancap\romannumeral #1@}
\makeatother

\newcommand{\sci}{Science}

\graphicspath{{./}{figures/}}

\received{\today}
\revised{tomorrow}
\accepted{the day after tomorrow}

\submitjournal{ApJ Letters}

\shorttitle{Shapes of HI-bearing dwarfs}
\shortauthors{Rong et al.}

\begin{document}

\title{Intrinsic Morphology of The Stellar Components in HI-bearing Dwarf Galaxies and The Dependence on Mass}

\correspondingauthor{Yu Rong}
\email{rongyua@ustc.edu.cn}

\author{Yu Rong}
\affiliation{Department of Astronomy, CAS Key Laboratory for Research in Galaxies and Cosmology, University of Science and Technology of China, Hefei 230026, P.R.China}
\affiliation{School of Astronomy and Space Science, University of Science and Technology of China, Hefei 230026, P.R. China}

\author{Min He}
\affiliation{National Astronomical Observatories, Chinese Academy of Sciences, Beijing 100012, China}

\author{Huijie Hu}
\affiliation{University of Chinese Academy of Sciences, Beijing 100049, China}
\affiliation{National Astronomical Observatories, Chinese Academy of Sciences, Beijing 100012, China}

\author{Hong-Xin Zhang}
\affiliation{Department of Astronomy, CAS Key Laboratory for Research in Galaxies and Cosmology, University of Science and Technology of China, Hefei 230026, P.R.China}
\affiliation{School of Astronomy and Space Science, University of Science and Technology of China, Hefei 230026, P.R. China}

\author{Hui-Yuan Wang}
\affiliation{Department of Astronomy, CAS Key Laboratory for Research in Galaxies and Cosmology, University of Science and Technology of China, Hefei 230026, P.R.China}
\affiliation{School of Astronomy and Space Science, University of Science and Technology of China, Hefei 230026, P.R. China}

\begin{abstract}
	
	The intrinsic morphology of stellar components within HI-bearing dwarf galaxies remains a topic of uncertainty. Leveraging the galaxy dataset derived from the cross-matched catalog of the Arecibo Legacy Fast Arecibo L-band Feed Array HI 21cm line survey and the Sloan Digital Sky Survey, we employ a Markov Chain Monte Carlo methodology and assume a triaxial model to scrutinize the inherent stellar distributions of these HI-bearing dwarf galaxies. Our analysis indicates a preference for oblate-triaxial models with $C<B\lesssim A$, indicative of thick stellar disks, characterizing the stellar components in these HI-bearing dwarfs with stellar masses ranging between $10^7\--10^{9.5}\ M_{\odot}$. The average thickness of the stellar components in HI-bearing dwarf galaxies approximates $C/A\sim 0.4$. Furthermore, we observe that the thickness of the stellar disks exhibits weak or negligible dependence on the stellar masses of HI-bearing galaxies.

\end{abstract}



\section{Introduction} \label{sec:1}

In empirical models of galaxy formation, the three-dimensional (3D) morphology of a galaxy arises from the intricate interplay among factors including galaxy mass, specific angular momentum, environmental influences, star formation history, and feedback \citep{Kaufmann07, Kado20, Rong20, Ball08, Lee13}. As a result, galaxy morphology can serve as a valuable indicator of these factors. Establishing the relationships between morphology and the diverse properties of galaxies is crucial for unraveling the complex mechanisms that govern galaxy formation and evolution.

High-resolution simulations and observations of high-redshift galaxies have revealed that galaxies accreting gas in the early universe often display clumpy star formation patterns, resulting in ``early-type'' morphologies \citep{Ma20, Nakazato24, Chen23, Fujimoto24}. Conversely, the formation of thin galactic discs occurs when gas accretion and star formation take place at lower redshifts \citep{Mo24, Forbes14, Lian24}. This distinction arises from the diverse modes of gas accretion during different epochs of the universe. In addition to the epoch of formation, galaxy morphology is influenced by both internal and external galactic properties.

For massive galaxies, their morphology is intricately entwined with their immediate environments \citep{vanderWel08, Padilla08, Rodriguez13}. Galaxies in denser regions typically exhibit more centrally concentrated morphologies, with a higher prevalence of elliptical galaxies observed in such environments \citep{Christlein05, Calvi12, Einasto14}. Moreover, galaxy morphology is affected by intrinsic factors such as stellar masses. At higher galaxy masses, the effects of dissipationless mergers become more prominent \citep{Oser10, Clauwens17}, leading to the formation of spheroids with significant asphericity at the high-mass end \citep{Holden12, Foster17, Krajnovic18, Li18}. Mergers also result in morphologies that are more centrally concentrated compared to their progenitor galaxies \citep{Martig12, Aumer14}. Additionally, galaxy morphology is intricately linked to the star formation history and stellar ages of the galaxy \citep{Christlein05, Guglielmo15}. Older stellar populations are typically associated with more centrally concentrated morphologies, while younger, actively star-forming galaxies tend to exhibit more diffuse morphologies \citep{Snyder15}.

Compared to massive galaxies, dwarf galaxies exhibit a stronger connection to their immediate surroundings. Dwarf galaxies located in denser environments are more vulnerable to gas loss, which can impact their morphological characteristics \citep[e.g.,][]{Tolstoy09, Rong17a}. Typically, dwarf galaxies are thicker than the rotating cold discs commonly seen in larger galaxies \citep{Sanchez-Janssen10, Roychowdhury13}, indicating that in these low-mass systems, dispersion support outweighs rotation \citep{Wheeler17}. This phenomenon suggests that feedback mechanisms play a crucial role in shaping the internal structure of low-mass centrals \citep{Kaufmann07, Governato10, Pontzen12, El-Badry16}. The dependence of dwarf galaxy morphology on stellar mass differs from that of massive galaxies \citep[e.g.,][]{Sanchez-Janssen19}, pointing towards distinct formation pathways for dwarfs \citep{Kormendy96}. Beyond environmental factors and mass, the chemical properties of dwarf galaxies, such as blue compact dwarfs, are also intertwined with their morphological characteristics \citep{Kunth00, Tolstoy09}. Moreover, the presence of nucleation has been identified as a factor influencing the morphology of dwarf galaxies \citep[e.g.,][]{Sanchez-Janssen19, Lisker07}. Examining the morphology and structure of dwarf galaxies offers valuable insights into their formation and evolutionary processes.

Numerous studies have delved into the morphologies of stellar components within massive galaxies \citep[e.g.,][]{Padilla08, Rodriguez13} and dwarf galaxies \citep[e.g.,][]{Sanchez-Janssen19, McConnachie12, Taylor17, Taylor18, Chen23}. However, the intrinsic morphology of stellar constituents in HI-bearing dwarf galaxies remains enigmatic. Unraveling the intrinsic morphology of HI-bearing galaxies not only illuminates the impact of internal feedback on shaping stellar distributions within these galaxies undergoing gas accretion and cooling but also holds paramount importance for precise estimations of their inclinations and halo masses \citep[e.g.,][]{Rong24}.

Fortunately, the Arecibo Legacy Fast Arecibo L-band Feed Array HI 21cm line survey (ALFALFA) \citep{Giovanelli05, Haynes18} has provided a substantial sample of HI-bearing dwarf galaxies, while the ongoing FAST HI survey is poised to deliver the most extensive HI-bearing dwarf galaxy dataset globally \citep{Zhang24}. These comprehensive surveys offer us the opportunity to statistically analyze the intrinsic morphologies of the stellar constituents in HI-bearing dwarfs by cross-referencing HI sources with optical data. In Section~\ref{sec:2}, we will present the samples under scrutiny in this study, selected from the ALFALFA dataset. Section~\ref{sec:3} will delve into investigating the intrinsic shapes of HI-bearing dwarf galaxies by employing a triaxial model and exploring the morphology's dependence on mass. Our findings will be discussed and summarized in Sections~\ref{sec:4} and \ref{sec:5}.

\section{HI-bearing galaxy sample in observation}\label{sec:2}

Our sample is derived from the cross-matched catalog of ALFALFA and Sloan Digital Sky Survey (SDSS DR12; Alam et al. 2015). ALFALFA represents an extensive extragalactic HI survey covering a vast area of approximately 6,600 deg$^2$ at high Galactic latitudes. The ALFALFA collaboration has released a fully comprehensive catalog  \citep[$\alpha.$100;][]{Haynes18} containing around 31,500 sources with radial velocities below 18,000 km s$^{-1}$, providing rich information for each source, including the signal-to-noise ratio (SNR) of the HI spectrum, cosmological distance, 50\% peak width of the HI line ($W_{50}$), and the HI mass ($M_{\rm{HI}}$), among other properties.

We have excluded sources with compromised photometric data or significant merging/interacting galaxy companions, as these companions have the potential to significantly perturb the morphology of the target galaxies. Additionally, we have omitted the inclusion of `dark galaxies' lacking optical counterparts or exhibiting exceedingly faint optical signatures. For each HI-bearing galaxy with an optical counterpart, we have meticulously utilized the \textsc{SExtractor} software \citep{Bertin1996} to precisely measure its $g$ and $i$-band `mag\_auto' magnitudes, effective radius $R_{\star,\rm{e}}$, and apparent axis ratio following the methodologies outlined by \cite{Du15}. Subsequently, the newly obtained photometric magnitudes are cross-checked against the photometry (mag\_cModel) provided by the SDSS pipeline.

Our analysis has revealed that while the SDSS magnitudes adequately represent the majority of galaxies, a small subset (e.g., AGC~1747, AGC~111801, and AGC~210221) exhibit inaccuracies in their `cModel' magnitudes, showing significant discrepancies from the newly acquired magnitudes, sometimes exceeding $\Delta i\gtrsim 2$~mag. In cases where the magnitude differences $\Delta i>0.75$~mag (i.e., twice the luminosities), we have meticulously applied \textsc{Galfit} \citep{Peng02} to reassess their magnitudes, ensuring a robust determination of the magnitude for each galaxy.

To determine the dust extinction-corrected absolute magnitudes $M_g$ and $M_i$ for galaxies exhibiting diverse axis ratios and colors, we employ the methodology elucidated by \cite{Durbala20} to mitigate the impact of internal dust extinctions within these galactic systems. Additionally, we leverage the $i$-band absolute magnitudes $M_i$ and $g-i$ colors to infer galactic stellar masses $M_\star$, utilizing the mass-to-light ratios with a formulation of $\log(M_{\star}/L_i) = 0.70(g-i)-0.68-0.057$ \citep{Taylor11}, where the term -0.057 represents a correction factor accommodating a Kroupa initial mass function \citep{Kroupa02,Herrmann16}. The stellar masses estimated from colors exhibit a typical uncertainty of approximately $\sim 0.1$~dex and have demonstrated strong agreement with the values documented in the GALEX-SDSS-WISE Legacy Catalog 2 \citep{Salim16,Salim18}, derived from ultraviolet-optical-infrared Spectral Energy Distribution (SED) fitting.

\begin{figure}[!]
	\centering{ \includegraphics[width=1.0\columnwidth]{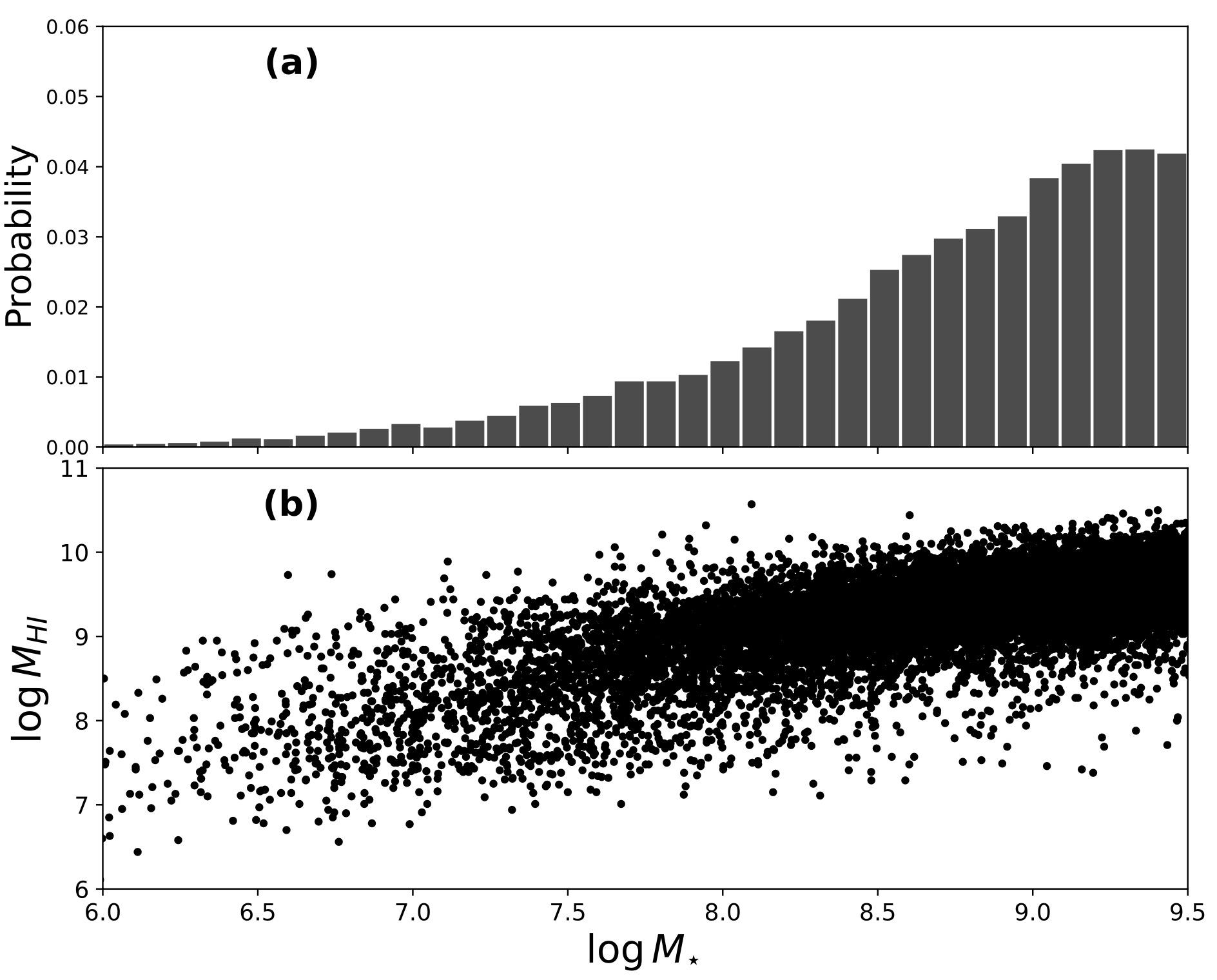} }
 	\vspace{-4mm}
 	 \caption{Panel~a: The stellar mass distribution of dwarf galaxies hosting HI in the ALFALFA survey. Panel~b: HI mass versus stellar mass diagram for these HI-bearing dwarf galaxies.
         }
	  \label{distribution}
\end{figure}

In this investigation, our objective is to elucidate the inherent morphology of the stellar constituents within dwarf galaxies, thereby excluding galaxies with stellar masses exceeding $10^{9.5}\ M_{\odot}$. The distributions of stellar mass and the correlation between the HI masses and stellar masses of HI-bearing dwarf galaxies are depicted in panels a and b of Fig.~\ref{distribution}, respectively. Given the significant influence of stellar masses on the intrinsic stellar configurations of galaxies \citep[e.g.,][]{Rong20,Sanchez-Janssen19,Roychowdhury13}, we categorize the HI-bearing galaxies into distinct stellar mass bins, encompassing $10^7<M_{\star}/M_{\odot}<10^{8}$, $10^8<M_{\star}/M_{\odot}<10^{8.5}$, $10^{8.5}<M_{\star}/M_{\odot}<10^{9}$, and $10^9<M_{\star}/M_{\odot}<10^{9.5}$, respectively. Subsequently, we compute the average intrinsic stellar configurations within each subset.

\section{Intrinsic morphology of HI-bearing dwarf galaxies}\label{sec:3}

\subsection{Modelling}\label{sec:3.1}

In this investigation, we employ a Markov Chain Monte Carlo (MCMC) methodology to probe the intrinsic morphologies of the stellar constituents within the HI-bearing galaxies. Drawing on the framework established by \cite{Rong20}, we undertake an analysis of the intrinsic morphology, offering a succinct overview of the methodological essentials herein, while directing interested readers to consult \cite{Sanchez-Janssen16} and \cite{Rong20} for a comprehensive exposition of the methodology.

Within this approach, the stellar constituents of galaxies within a specific subset (e.g., characterized by comparable stellar masses) are conceptualized as a suite of optically-thin triaxial ellipsoids. The 3D density distribution of the galaxy is delineated by a collection of aligned ellipsoids, each defined by a common ellipticity $E=1-C/A$ and a triaxiality $T = (A^2-B^2)/(A^2-C^2)$ \citep{Franx91}, where $A\geq B\geq C$ represent the intrinsic major, intermediate, and minor axes of the ellipsoid, respectively. 

Within a subsample of HI-bearing galaxies, it is posited that their ellipticity $E$ and $T$ adhere to Gaussian distributions characterized by mean values $(\langle E \rangle$, $\langle T \rangle)$ and standard deviations $(\sigma_E$, $\sigma_T)$. By considering the distribution of intrinsic axis ratios and the stochastic nature of viewing angles for the model galaxies, the apparent axis ratios $b/a$ distribution can be deduced through the projection of these ellipsoids \citep[see the Appendix of ][for detailed projection methodology]{Rong15a}. 

Consequently, the posterior probability density function (pdf) of the model parameters $(\langle E \rangle$, $\sigma_E$, $\langle T \rangle$, and $\sigma_T)$, emblematic of the intrinsic shapes of the stellar constituents within each HI-bearing galaxy subset, can be ascertained by embracing a Bayesian framework \citep[cf.][]{Sanchez-Janssen16}. It is assumed that the prior probabilities of $\langle E \rangle$ and $\langle T \rangle$ conform to uniform distributions in the interval 
 $[0,1]$, while $\sigma_E$ and $\sigma_T$ follow a scaling $p(\sigma)\propto \sigma^{-1}$. To explore the posterior distribution of the model parameters, we employ the {\textsc{emcee}} code \citep{Foreman-Mackey13} with 100 `walkers' and 5000 steps, ensuring sufficient convergence of the MCMC chains.

\begin{figure*}[!]
	\centering{ \includegraphics[width=1.0\textwidth]{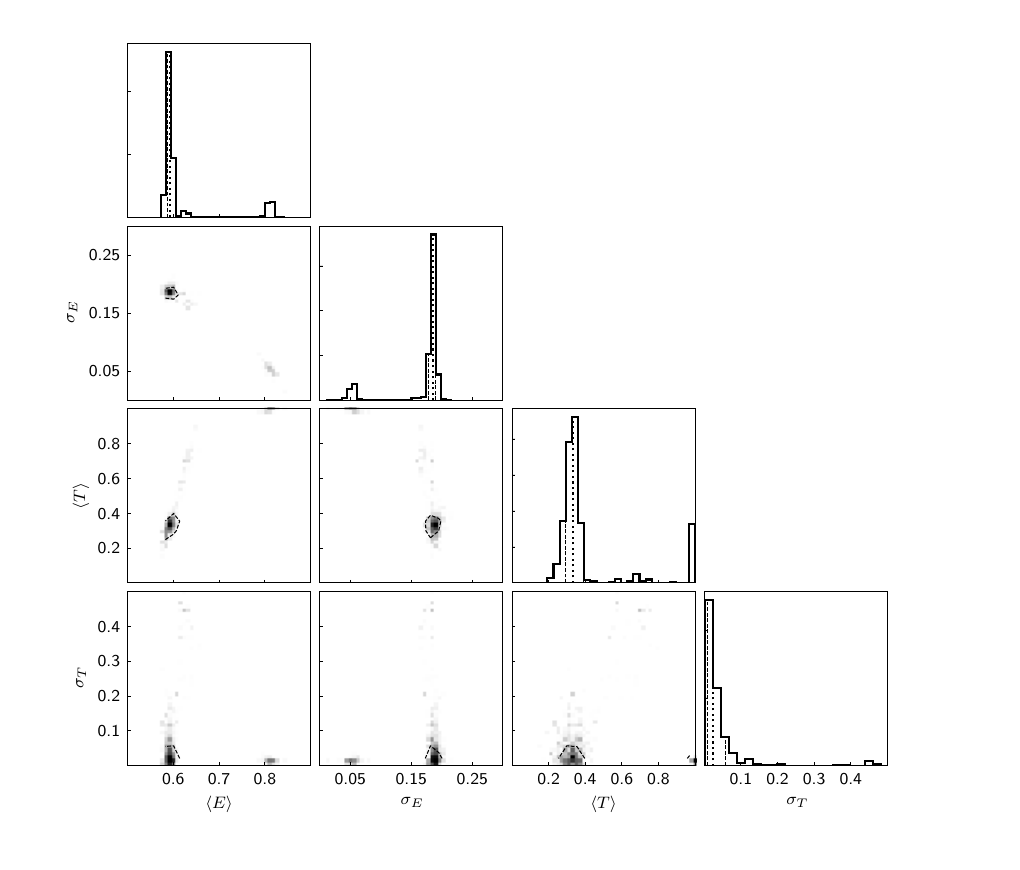} }
	\vspace{-4mm}
 	 \caption{Posterior probability density functions of $(\langle E \rangle$, $\sigma_E$, $\langle T \rangle$, and $\sigma_T)$ for HI-bearing dwarf galaxies with stellar masses of $10^{8.5}<M_{\star}/M_{\odot}<10^9$, selected from ALFALFA. The panels in the diagonal show the posterior pdfs for each of the parameters, marginalized over all the other ones. The grey scale in the non-diagonal panels shows the corresponding joint posterior pdfs. Contours enclose the regions that contain 68\% of the cumulative posterior probability. The dotted and dashed lines in the diagonal panels indicate the 50\% and 16\% and 84\% of the corresponding marginalized posteriors, respectively.
         }
	  \label{MC}
\end{figure*}


\subsection{Results}\label{sec:3.2}

We present a synthesis of the modeling outcomes for the subsamples of HI-bearing galaxies in Table~\ref{shape}, alongside the posterior distributions of $\langle E \rangle$, $\sigma_E$, $\langle T \rangle$, and $\sigma_T$ for a representative subsample characterized by a stellar mass range of $10^{8.5}<M_{\star}/M_{\odot}<10^9$ in Fig.~\ref{MC}. The posterior distributions of $\langle E \rangle$ and $\langle T \rangle$ consistently exhibit bimodal features, hinting at the potential presence of two distinct underlying configurations for the HI-hosting dwarf galaxies: the peak at $\langle E \rangle\sim 0.6$ and $\langle T \rangle\sim 0.3$ corresponds to an oblate-triaxial model ($C/B\ll B/A$), indicative of a ``thick disk morphology'', while the peak at $\langle E \rangle\sim 0.8$ and $\langle T \rangle\sim 1.0$ aligns with a predominantly prolate model ($C\simeq B\ll A$). Our findings suggest that, on a statistical basis, these HI-bearing dwarf galaxies are most likely to exhibit a thick-disk morphology, although the presence of a small subset displaying a prolate configuration cannot be entirely discounted.

\begin{figure*}
	\centerline{ \includegraphics[width=1.0\textwidth]{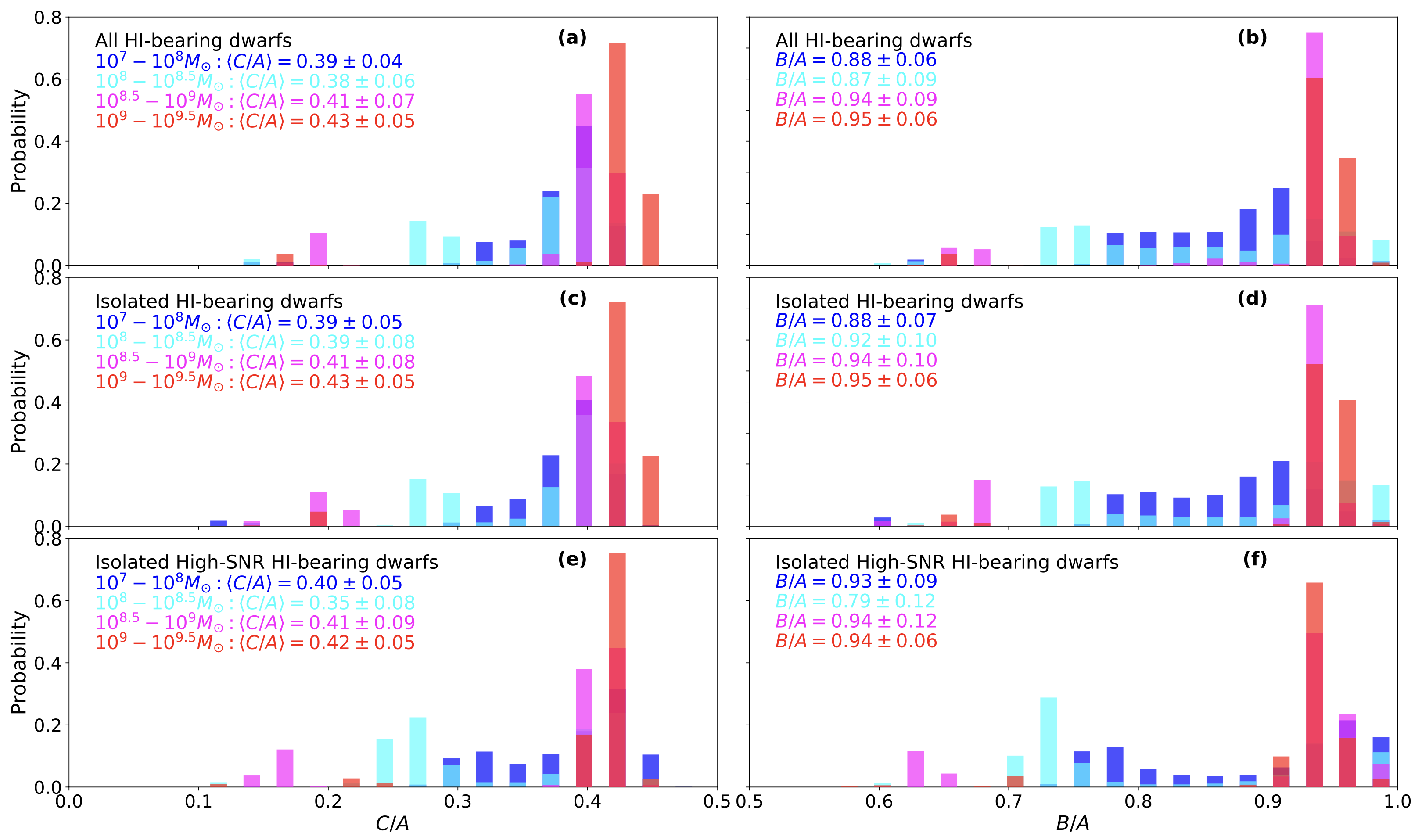} }
	\vspace{-4mm}
 	 \caption{The intrinsic axis ratios $\langle C/A \rangle$ (left panels) and $\langle B/A \rangle$ (right panels) of the HI-bearing dwarf galaxies, derived from the posterior probability density functions of $\langle E \rangle$ and $\langle T \rangle$. The results for the entire dwarf sample, isolated dwarf sample, and isolated high-SNR dwarf sample are presented in the upper, middle, and lower panels, respectively. Stellar mass bins of $10^7<M_{\star}/M_{\odot}<10^{8}$, $10^8<M_{\star}/M_{\odot}<10^{8.5}$, $10^{8.5}<M_{\star}/M_{\odot}<10^{9}$, and $10^9<M_{\star}/M_{\odot}<10^{9.5}$ are color-coded in blue, cyan, magenta, and red, respectively.
         }
	  \label{CA_comp}
\end{figure*}

The intrinsic average axis ratios, $\langle C/A \rangle$ and $\langle B/A \rangle$, are subsequently derived from the parameters $\langle E \rangle$ and $\langle T \rangle$. The morphological characteristics exhibit a subtle dependence on galactic stellar mass, as illustrated in Fig.~\ref{CA_comp}. Notably, the average disk thickness, quantified by $\langle C/A \rangle$, shows a marginal increase with rising stellar masses on a statistical basis, while the ratio $\langle B/A \rangle$ also displays a slight elevation with increasing stellar mass{\footnote{Given the challenges in constraining triaxiality distributions solely from photometric data \citep{Binggeli80}, our analysis focuses on comparing galaxy flattenings ($\langle E \rangle$) and thicknesses ($C/A$).}}. However, the magnitude of the rise in $\langle C/A \rangle$ with stellar mass is modest, hovering around $\langle C/A \rangle \sim 0.4$. Considering the uncertainties in the $\langle C/A \rangle$ ratios, it can be argued that the disk thickness does not exhibit significant variations with galactic mass.

\begin{table*}[!]  \centering 
\begin{tabular}{@{}c|ccccccc@{}}
\hline
\hline
Sample & Stellar mass range & $N$ & $\langle E \rangle$ & $\sigma_E$ & $\langle T \rangle$ & $\sigma_T$ & $\langle A:B:C \rangle$ \\
\hline
 & $\log M_{\star}\sim [7,8]$ & 1937 & $0.61_{-0.02}^{+0.01}$ & $0.14_{-0.01}^{+0.01}$ & $0.60_{-0.16}^{+0.28}$ & $0.20_{-0.18}^{+0.20}$ & $1:0.88:0.39$ \\
 & $\log M_{\star}\sim [8,8.5]$ & 2463 & $0.62_{-0.03}^{+0.01}$ & $0.15_{-0.04}^{+0.01}$ & $0.64_{-0.40}^{+0.31}$ & $0.32_{-0.29}^{+0.17}$ & $1:0.87:0.38$ \\
All & $\log M_{\star}\sim [8.5,9]$ & 4304 & $0.59_{-0.01}^{+0.01}$ & $0.19_{-0.01}^{+0.01}$ & $0.33_{-0.04}^{+0.06}$ & $0.02_{-0.01}^{+0.04}$ & $1:0.94:0.41$ \\
 & $\log M_{\star}\sim [9,9.5]$ & 5929 & $0.57_{-0.01}^{+0.01}$ & $0.21_{-0.01}^{+0.01}$ & $0.30_{-0.04}^{+0.04}$ & $0.03_{-0.02}^{+0.04}$ & $1:0.95:0.43$ \\
\hline

 & $\log M_{\star}\sim [7,8]$ & 1538 & $0.61_{-0.02}^{+0.01}$ & $0.14_{-0.01}^{+0.01}$ & $0.59_{-0.20}^{+0.29}$ & $0.22_{-0.20}^{+0.20}$ & $1:0.88:0.39$ \\
 & $\log M_{\star}\sim [8,8.5]$ & 2036 & $0.61_{-0.02}^{+0.01}$ & $0.15_{-0.05}^{+0.01}$ & $0.41_{-0.24}^{+0.53}$ & $0.26_{-0.23}^{+0.21}$ & $1:0.92:0.39$ \\
Isolated & $\log M_{\star}\sim [8.5,9]$ & 3648 & $0.59_{-0.01}^{+0.01}$ & $0.18_{-0.12}^{+0.01}$ & $0.35_{-0.05}^{+0.63}$ & $0.02_{-0.01}^{+0.05}$ & $1:0.94:0.41$ \\
 & $\log M_{\star}\sim [9,9.5]$ & 5090 & $0.57_{-0.01}^{+0.01}$ & $0.21_{-0.01}^{+0.01}$ & $0.30_{-0.04}^{+0.05}$ & $-0.02_{-0.02}^{+0.05}$ & $1:0.95:0.43$ \\
\hline

 & $\log M_{\star}\sim [7,8]$ & 678 & $0.60_{-0.03}^{+0.01}$ & $0.14_{-0.03}^{+0.01}$ & $0.39_{-0.25}^{+0.53}$ & $0.14_{-0.13}^{+0.25}$ & $1:0.93:0.40$ \\
 & $\log M_{\star}\sim [8,8.5]$ & 784 & $0.65_{-0.07}^{+0.01}$ & $0.15_{-0.05}^{+0.03}$ & $0.90_{-0.71}^{+0.07}$ & $0.06_{-0.04}^{+0.24}$ & $1:0.79:0.35$ \\
Isolated, high SNR & $\log M_{\star}\sim [8.5,9]$ & 1299 & $0.59_{-0.01}^{+0.01}$ & $0.20_{-0.15}^{+0.01}$ & $0.32_{-0.09}^{+0.67}$ & $0.02_{-0.01}^{+0.12}$ & $1:0.94:0.41$ \\
 & $\log M_{\star}\sim [9,9.5]$ & 1713 & $0.58_{-0.01}^{+0.01}$ & $0.21_{-0.01}^{+0.01}$ & $0.34_{-0.06}^{+0.07}$ & $0.03_{-0.02}^{+0.14}$ & $1:0.94:0.42$ \\
\hline
\hline
\end{tabular}
\caption{MCMC results for the intrinsic morphology analysis of distinct subsamples of HI-bearing dwarf galaxies.
Column (1): The HI-bearing galaxy sample. The intrinsic morphology of the entire ALFALFA dataset (All), the isolated sample with $R/R_{\rm{vir}}>3$, and the high signal-to-noise ratio (SNR) sample with HI SNR$>15$ are examined in this investigation. The numbers in parentheses indicate the corresponding sample sizes. Column (2): Various stellar masses within the HI-bearing galaxy subsample. Column (3): The counts of HI-bearing galaxies in the subsamples used for MCMC analysis. Columns (4)-(7): Mean values and standard deviations of ellipticity and triaxiality distributions, represented as $\langle E \rangle$, $\sigma_E$, $\langle T \rangle$, and $\sigma_T$. The $1\sigma$ uncertainty for each parameter is also reported. Column (8): Average intrinsic axis ratios $\langle A:B:C \rangle$.}
\label{shape}
\end{table*}

It is worth noting that a subset of HI-bearing dwarf galaxies are located in close proximity to galaxy clusters or groups, where their morphologies may have been influenced by tidal effects from the dense environments, potentially altering their intrinsic stellar distributions \citep{Moore96,Larson80,Kawata08,Bekki09,Mayer07,Read05}. Therefore, we further exclude these samples situated near high-density regions and reevaluate the morphologies of HI-bearing dwarf galaxies using the remaining isolated samples. To ascertain the spatial contexts of these HI-bearing dwarf galaxies, we leverage the galaxy group and cluster inventory curated by \cite{Saulder16}. This catalog is meticulously assembled utilizing data from the SDSS DR12 and 2MASS Redshift Survey \citep{Huchra12}, employing the friends-of-friends group identification algorithm. Notably, the investigation by \cite{Saulder16} rigorously addresses diverse observational biases, including the Malmquist bias and the `Fingers of God' effect. Galaxies are designated as non-isolated if they lie within a distance of three times the virial radius of a galaxy group or cluster (i.e., $R/R_{\rm{vir}}\leq 3$). Conversely, galaxies failing to meet this criterion are classified as isolated.

As illustrated in panels c and d of Fig.~\ref{CA_comp} and Table~\ref{shape}, the ratios of $\langle C/A \rangle$ and $\langle B/A \rangle$ for the isolated HI-bearing galaxies exhibit no noteworthy distinctions compared to the entire sample. This observation can be attributed to the limited presence of non-isolated galaxies within our HI-bearing dataset.

\section{Discussion}\label{sec:4}

\subsection{Sample completeness test}

First and foremost, it is pertinent to highlight that our sample does not include a subset of ALFALFA galaxies known as ``dark galaxies'', which lack optical counterparts or display extremely faint optical signatures \citep{Disney76,Janowiecki15}. These dark galaxies have been identified in previous studies as prone to tidal interactions \citep{Roman21,Duc08}, rendering them non-equilibrium systems that should be excluded from our analysis. However, given their limited presence, their exclusion does not impact the statistical investigations into the intrinsic morphology of HI-bearing galaxies.

Secondly, our exploration of the stellar distributions within these HI-bearing galaxies is markedly influenced by the completeness of a subset of dwarf galaxies, which is bound by the detection thresholds of the ALFALFA telescope. For instance, in scenarios where two galaxies possess comparable HI masses, the galaxy with a larger HI line width, denoted as $W_{\rm{50}}$, may exhibit a lower SNR in its HI emission compared to the galaxy with a smaller $W_{\rm{50}}$, potentially rendering it undetectable by the Arecibo. To assess the potential biases in our morphological estimations, we impose a stringent SNR threshold of SNR$>15$ and investigate whether the ratios of $\langle C/A \rangle$ and $\langle B/A \rangle$ exhibit any variations with respect to this threshold. As delineated in Table~\ref{shape} and illustrated in Fig.~\ref{CA_comp}, the 3D axis ratios for the entire sample and those exceeding the SNR threshold of 15 demonstrate no discernible biases, implying that the impact of the SNR cutoff is negligible.



\subsection{Comparison with previous studies for morphology}

Despite the scarcity of research on the intrinsic morphologies of stellar components within HI-bearing galaxies, significant efforts have been devoted to investigating the morphologies of stellar constituents in dwarf galaxies located in relatively high-density environments \citep[e.g.,][]{Sanchez-Janssen19,Rong20,McConnachie12,Taylor17,Taylor18,Roychowdhury13,Salomon15,Lisker07,Ryden94,Ichikawa89}, characterized by subdued star-formation rates, as well as in massive galaxies \citep[e.g.,][]{Weijmans14,Padilla08,Rodriguez13,Holden12}. Notably, studies such as that by \cite{Sanchez-Janssen19} have revealed that the morphologies of dE/dSphs within the Coma, Virgo, and Fornax clusters are contingent upon galactic stellar masses. Additionally, research by \cite{McConnachie12} and \cite{Taylor17,Taylor18} have delved into the intrinsic morphologies of dEs/dSphs within the Local Group and the Centaurus~A galaxy group, respectively. In light of these findings, the intrinsic thickness of the dEs/dSphs located in proximate galaxy clusters and groups decreases with increasing stellar mass.

These results contrast with our findings, which indicate a weak or non-existent thickness dependence on stellar mass. This discrepancy may be attributed to the absence of extremely low-mass dwarf galaxies with $M_{\star}<10^7\ M_{\odot}$ in our sample of HI-bearing dwarf galaxies. The relatively high mass of our galaxy sample may contribute to the lack of significant thickening of disks with decreasing galaxy mass observed. It is also possible that our HI-bearing dwarf galaxies at low redshifts are likely undergoing recent gas accretion, which may occur more prominently along specific directions with specific angular momentum, leading to the formation of gas disks. The stars formed from the newly accreted gas are distributed on a relatively flattened, thick disk. The light emitted by these new stars contaminates the original, old stellar distribution, resulting in stellar disks that exhibit a nearly uniform thickness regardless of mass. In contrast, the formation of old stars from gas accretion in the early Universe may lead to less flattened structures, as gas accretion during high-redshift periods is believed to be more chaotic and isotropic in nature \citep[e.g.,][]{Keres09,Ma20, Nakazato24, Chen23, Fujimoto24}.

Furthermore, the star formation activity in dwarf galaxies is expected to occur in intermittent bursts across various redshifts \citep{Muratov15}, with supernova-triggered outflows exerting pressure on gas and heating stellar orbits \citep{Pontzen12,El-Badry16,Teyssier13,Governato10}. Therefore, we propose that more massive HI-bearing dwarf galaxies have undergone a higher frequency of starbursts or a greater volume of star formation, leading to an increased number of supernova events. The cumulative effect of multiple supernova explosions induces vigorous and recurrent perturbations in the gravitational potential of dwarfs, facilitating energy transfer to collisionless components like dark matter and stars. Consequently, it is likely that higher-mass HI-bearing dwarfs have contributed more energy to enhance the random motions of stars, thereby mitigating the effect of thinning thickness with increasing mass.

It is also noteworthy that the observed thickness dependence on mass among dEs/dSphs in high-density environments likely reflects the influence of tidal forces, as lower-mass galaxies possess shallower gravitational wells and are more susceptible to tidal perturbations from massive galaxies \citep[][]{Moore96,Mayer01,Mayer07,Mastropietro05,Smith15,Kazantzidis11}, which can effectively enhance the structural thickness of dwarf galaxies. In contrast, the distinctive features of HI-bearing dwarf galaxies suggest that tidal interactions play a minimal role in shaping their morphologies, as they appear to be largely unaffected by their environment.

\section{Summary}\label{sec:5}

We utilized the $g$-band apparent axis ratios of dwarf galaxies with stellar masses ranging from $10^7< M_{\star}/M_{\odot} <10^{9.5}$ in the ALFALFA dataset and employed a Markov Chain Monte Carlo methodology, assuming a triaxial model, to investigate the intrinsic morphologies of the stellar components within these HI-bearing dwarf galaxies. Our analysis reveals that these HI-bearing dwarfs likely exhibit thick-disk morphologies, characterized by an average thickness of approximately  $\langle C/A \rangle \sim 0.4$. Remarkably, the statistical analysis indicates that the average thickness $\langle C/A \rangle$ exhibits weak or negligible dependence on the stellar masses of HI-bearing dwarf galaxies.
\vspace{10mm}

\begin{acknowledgments}

Y.R. acknowledges supports from the NSFC grant 12273037, and the CAS Pioneer Hundred Talents Program (Category B), as well as the USTC Research Funds of the Double First-Class Initiative. H.X.Z. acknowledges support from the NSFC grant 11421303. H.Y.W. is supported by CAS Project for Young Scientists in Basic Research, Grant No. YSBR-062, and the NSFC grant 12192224.

\end{acknowledgments}



\end{document}